\documentstyle[a4,12pt,epsfig]{article}

\newcommand{\fslash}{\hspace{-0.26cm} /}
\begin{document}
\title{Combination of Heavy Quark with Partons from Quark-Gluon Matter: a Scaling Probe}


\author{Shi-Yuan   Li\\
 {\small \it Institute of Particle Physics, Huazhong  Normal University, Wuhan, 430079}\\
 {\small \it  and   Department of Physics, Shandong University, Jinan, 250100, PR China}}

\maketitle

\begin{abstract}
%
In relativistic heavy ion collisions,
%
%
 the cross section of heavy hadron production  via the  combination of a heavy quark with a light one from the quark-gluon matter can be factorized. It is the convolution of   twist-4 combination matrix elements, the parameters corresponding to the  parton distributions of the quark-gluon matter, as well as the hard partonic  cross section of heavy quark production calculable in  PQCD. 
These parton distributions and  combination matrix elements are functions of a scaling variable which is the momentum fraction of the heavy quark w.r.t. the heavy hadron.  In the same factorization framework, the combination matrix elements appear in other `simpler'  processes and can be extracted. Taking them  as inputs, comparing with data from  RHIC and (future) LHC, we can get the parton distributions of the quark-gluon matter just as the similar way we get those of nucleon, pion, or photon, etc.. 
\end{abstract}

\section{Introduction}
\label{sec:intro}

One of the main purposes  of (ultra-)relativistic  heavy ion collision experiments,  e.g., those in RHIC at  BNL and LHC at CERN, is to produce and study a deconfined phase of quarks and gluons, usually referred to as  Quark-Gluon Plasma (QGP), under the extreme conditions at
high temperature and/or high  density.  QGP is an important prediction of  Quantum Chromodynamics (QCD) and is   supposed  been existing in the early universe shortly after the Big Bang  \cite{preqgp}. In the past few years, the hot dense  quark-gluon matter has been produced at RHIC,  while experimentalists  are still working on   more  signals \cite{signal}. At such a stage,  aware of  the particular  complexity of the experiments, one can be more conscious than ever that to get more knowledge of QCD at extreme conditions and/or to `simulate' the state of the universe at early time on collider, one should go further than just to `discover' QGP.  That is, to carefully  and {\bf quantitatively}  measure it. In fact, it is widely agreed among experimentalists and theorists that  more detailed measurements  are needed before drawing a definite conclusion on what is produced at RHIC
\cite{signal, nature}.
There have been many marvelous probes suggested, which have played  the  main r\^{o}le  in the study of  the hot dense quark-gluon matter at RHIC \cite{signal, jacobwang}. 
On the other hand, new methods  with unambiguous quantitative relation between the final state particles  and the partons of QGP (or other states of quark-gluon matter produced in the collision) are still worthy to be looked for. 
  In this note, we study the possibility of measuring the Quark-Gluon Matter (QGM) via the heavy hadron  produced by   a heavy quark (charm or bottom) created in the hard collision and combined with the partons from the QGM. The data from  RHIC themselves are the suggestion for this idea.

Recently, some  `unexpected' phenomena on hadron production in Au-Au collision at RHIC (\cite{signal} and refs. therein), e.g.,  different $p_\perp$ suppression between mesons and baryons and the `$v_2$ scaling', challenge the `energy loss + fragmentation' picture of hadron production and are well explained by various combination models \cite{fries, hwa, ko}. 
Besides, these models  provide a clue to measure  the partons in the QGP/QGM   because the production rates of the hadrons are  proportional (via convolution) to the distribution function of  the quarks (gluons \footnote{The phenomenological models cited above only  consider the valence quarks of a meson or baryon  now. }) in the QGP/QGM. Hence one can derive, or at least infer the `parton distributions' from final hadrons. The work of \cite{hwa} is an example. The authors extract the quark distribution from the pion data within the framework of their model, and predict or explain  the production of the others. 


The quark (re)combination models   are of a long history. They can date back to three  decades ago. From then on, different kinds of quark (re)combination models  are presented in application to hadron production in various high energy processes\cite{annisovich73el}.
Therefore it is not surprising that several different combination models are suggested to explain the RHIC data as cited above and new works are emerging \cite{xieshao}.
 Unfortunately, such a condition puts forward  the problems and uncertainties:  Both the distributions and combination rules of these models are different from each other,  so one is difficult to gain the unambiguous knowledge on the content of the QGM. In other words, the `parton distribution' inferred in this way is model-dependent. If one wants to `measure' the parton distribution of QGP/QGM in the combination production processes, one has to fix the unique  `combination rules'. This seems beyond approach  for light hadrons now.

However, the combination of a heavy quark (charm or bottom) with light ones from the  QGM can provide  more clear information, and is free of model dependence. The main points are: 1) The combination production can be   factorized. The cross section formula (e.g., that of  the inclusive  process $A+B \to M_Q +X$, here $A$, $B$, $M_Q$ denote the colliding nuclei and the produced heavy  meson, respectively) is the convolution of the hard sub-cross section of the heavy quark production ($A+B \to Q \bar{Q}$),  the combination matrix elements and the parameters corresponding to the  parton distributions of the QGM. The combination matrix element and the parton distribution of the QGM are expectation values of field operators on certain particle states, so they are model-independent and process-independent. 
2) The spectra of heavy quarks can be calculated by PQCD (now full NLO corrections are available and tested in Tevatron \cite{NLOte}.). 3) Within the same factorization scheme, the combination matrix elements which describe the probability of a heavy quark and a light one of specified momenta to form a heavy hadron,   appear in other  more `simple and clean' processes, such as $e^+ e^-$ annihilation, DIS and hadronic collisions, so can be extracted for the experiments. 
From the above points, it is clear that in the cross section formula, taking Point 3) as inputs, comparing with heavy ion collision data, we can get the parton distribution of QGM.





 A simple comparison of the `combination probe' with the hard probe of the energy loss/jet quenching  can help to understand the idea in this note. We do not expect to find advantages or disadvantages,  but we find that they are complimentary for each other.
 Let's first see  the electroweak  probe (photon, $W^{+-}$, $Z^0$) on nucleons,
in which case, the  single photon (or other gauge bosons) approximation is enough. Imagining that if on the contrary,  we have to sum the electroweak  coupling to all  orders, i.e.,  multi-interaction between the electron (neutrino)  and the proton by exchanging multiple gauge bosons is important, we lose  the unambiguous definition of  $ W_{\mu \nu}$. This is the case for a hard jet  in QGM,  where multiple soft strong interactions between the hard parton and the QGM have to be taken into account. So, it is hardly possible for the hard jet to play the same r\^{o}le of the electron (neutrino).  However, for the combination process of a  heavy quark with the  light one from the QGM, 
 the cross section is proportional to   the parton distribution function of the QGM.  
It is more attractive in the results of this note that, for certain $Q$, which is the CMS energy of the partonic process of heavy quark production, the  heavy meson of momentum $K$ just comes from the heavy quark with momentum fraction $z=\frac{Q^2}{2K \cdot q}$ \footnote{This is exact for the $2 \to 2$ partonic process. For  final states with more partons, the relation is kept as long as we take $Q$ as the invariant mass of the heavy quark pair. As we shall see in Section \ref{sec:formucom}, this relation comes from the on shell condition of the `freely-fragmenting' quark and the 4-momentum conservation for the quark pair system, which are also the origin of the Bjorken scaling.}, hence probes the light quark in QGM with momentum $k=(1-z)K$. This is very similar to the Bjorken scaling in DIS.  
 
Heavy quarks are mostly produced from the initial parton scattering (in this note we do not discuss the possibility of creation of heavy quark pairs from the QCD vacuum fluctuation  at very high temperature), they evolve in space-time and interact  with the QGM.  The measurement of jet quenching can give important information of the  space-time structure of the QGM, e.g., the correlation length or the space scale of the medium,  $L$ \cite{dong}. On the other hand, the formulae in this note are  the result of the integration on the  whole space-time, so they are only sensitive to the local phase space (momentum) distributions of the partons.
 In fact, from the discussions in the following sections,  we can see that the   induced radiation  can be  included into the  evolution  of the partons or the matrix elements (`$z$ scaling violation'). So the energy loss and combination are both sub-processes in the heavy hadron production, and both are necessary in prediction of the experimental data. This note only concentrate on the combination mechanism.




 The outline of this note is: Section 2 introduces the derivation/factorization of the  cross section, taking the case of heavy meson as an example . In section 3 we show that  the combination matrix elements appear in and  can be extracted from other simpler  processes.
Section 4 gives the numerical estimation of the combination process cross section with the input of assumed  values of the quark distribution  of  `QGP'   and  the recombination matrix elements (since they are still beyond available from experiments now), only to demonstrate the feasibility of the  formulae.
  Section  5 is for conclusions and discussions. 
 The combination of heavy quark with gluons and heavy baryon production  are left for later works.


\section{Combination of heavy quark  with light one from QGM}
\label{sec:formucom}

 In the following of this note, we take $A+B \to \bar D + X$ as an  example. Here $\bar D$ refers to  any  anti-charm meson.   We only consider the contribution of $\bar c+q$ to $\bar D$, other possible combination processes  such as $\bar c+g$ to $\bar D$ are assumed negligible.   %
%
%
%
%
%
     The $X$ includes the associated produced c quark and all the other particles from the nucleus-nucleus $A$, $B$  interaction. 
In this section, we  derive the  inclusive invariant  differential  
 production number 
$2E \frac{d {\cal N }_C^{AB}}{d^3 \bf K}$ of $\bar D$. The  subscribe $C$ of   ${\cal N}$  denotes  the combination process. $(E, \bf {K})$ is the 4-momentum of  $\bar D$.  The production of charm mesons can be treated in the same way as the anti-charm mesons.

To describe the light quark from QGM, we should find ways to represent the `external particle source'.  Similar as  the works on energy loss \cite{dong, wanggy}, we employ  an external field to  describe  the interaction with QGM. 
 We can see in the following, this external field,  together with quark field operators, appears in the matrix element corresponding  to  the quark distribution  in the  QGM.  Same as \cite{qiuvitev},  we choose  the external field a vector $V^{\mu}_{ext}$ proportional to $n^{\mu}=(0, 1, {\bf 0}_{\perp})$. This can be understood as gauge fixing and is easy to factorize the Dirac indices. We would emphasize here that the external field can present hot dense quark gluon matter or QGP, as well as cold quark matter (nuclei), as in \cite{qiuvitev}. If one calculates the matrix elements by, e.g., lattice QCD, one should identify the concrete form of the external field and discuss the details of the differences between  hot and cold QGM. However, here we only use it in the formal expression of the matrix elements, which is to be probed in experiments. So $V_{ext}$ is just a `symbol' here.


From the above discussions, the interaction Hamiltonian for quark and gluon fields is extended as:
\begin{equation}
\label{intham}
H_I=\Psi ({  A \hspace{-0.26cm}  \slash}+ { V \fslash}) \Psi
\end{equation}

 Here $A$ is the normal gluon field and $V$  is the 
external field. 
The strong  coupling constant $g_s$ is absorbed into the gauge fields. In this note, wherever we write the gauge field obviously, we always adopt  this convention.  One can easily see  that this is the same  interaction Hamiltonian in describing  the energy loss in QGM by induced radiation, which has been discussed comprehensively in previous works, e.g., \cite{dong,wanggy}. In this note, we only concentrate on the combination. The jet quenching processes  in our frame work can be taken as radiation corrections (i.e., to consider higher orders of the perturbative expansion of the S-matrix)  in the QGM environment and will  be treated as the evolution of the  matrix elements (z scaling violation) in later works.


If  we assume that the distribution functions of the  intrinsic heavy flavours in the initial  nuclei are vanishing, the lowest order contribution for the recombination process comes from $O(g_s^3)$ in the perturbative  expansion of S-matrix:

\begin{eqnarray}
\label{smatrics}
 S^{(3)}& = &\frac{(-i)^3}{3 !}C^2_3 \int d^4 x_1 d^4 x_2 d^4 x_3 {\bf T} \bar \Psi(x_1)  A\fslash (x_1) \Psi(x_1)  \nonumber \\
&\times &\bar \Psi(x_2)  A\fslash (x_2) \Psi(x_2) \bar \Psi(x_3)  V\fslash (x_3) \Psi(x_3),    
\end{eqnarray}
where summation on colour and flavour indices as well as the indices in spinor space are indicated.
 
Now we  take the annihilation partonic process $q \bar q \to c \bar c$ as an example to illustrate the derivation, 
 while the total result can be obtained by summing all kinds of partonic processes.  Let the corresponding terms of the  Wick expansion
act on the initial nuclear state$|A B> $ (We will discuss the Glauber geometrical formulae \cite{glauber} in the end of this section.) and final state $| \bar D X>$, employing the space-time translation invariance, we can isolate the $\delta$ function corresponding to  the total energy-momentum conservation and get the T-matrix element. The  cross section  is, then

\begin{eqnarray}
\label{jiemian1}
\sigma&= & \frac{(4 \pi \alpha_s)^2}{4 F}  \sum \limits_{\bar D X}  \int d^4x_1 d^4x_2 d^4x_3 d^4x_4 d^4x_5  \int{ \frac{d^4 q}{(2 \pi)^4} \frac{d^4 q'}{(2\pi)^4}} \frac{1}{q^2 q'^2} e^{-iq(x_1-x_2)} e^{iq'(x_4-x_5)} \nonumber \\
 & \times &
< A B| \bar \psi(x_4) T^{c_4} \gamma^{\mu_4} \psi(x_4)  \bar \psi(x_3) \not V(x_3) \psi(x_3) \bar \Psi(x_5) T^{c_4} \gamma_{\mu_4}  \Psi(x_5)|\bar D X> \nonumber \\
 &\times & 
< \bar D X|  \bar \Psi(x_2) T^{c_1} \gamma_{\mu_1} \Psi(x_2)  \bar \psi(0) \not V(0) \psi(0) \bar \psi(x_1) T^{c_1} \gamma^{\mu_1}  \psi(x_1)|  A B>.
\end{eqnarray}

In the above equation, $T^c$ is one half of the Gell-Mann Matrix, $4 F$ represents the incident  flux factor and discrete quantum number average.  Summation on repeated indices is indicated. The capital  $\Psi$ is for the heavy quark field and the lower case $\psi$ for  light quark fields.

The following thing is to factorize the cross section in the framework of collinear factorization. The  cross section can be written as

\begin{equation}
\label{jiemian2}
\sigma=\frac{(4\pi \alpha_s)^2 C}{4 F} \int \frac{d^4 q}{(2\pi)^4} W^{\mu \nu} \frac{1}{q^4} D_{\mu \nu},
\end{equation}

with
 \begin{eqnarray}
 \label{wmunu}
 W^{\mu \nu}& = & \int d^4 x_4 e^{-iqx_4} < A | \bar \psi^{\alpha_{4}} (x_4) \psi^{\beta_{1}}(0)| A> <B| \psi^{\beta_4}(x_4) \bar \psi^{\alpha_1}(0)|B> \nonumber \\
&\times &  \gamma ^{\mu}_{\alpha_4 \beta_4} \gamma^{\nu}_{\alpha_1 \beta_1} + ( A  \leftrightarrow B).
\end{eqnarray}
This is the same $W^{\mu \nu}$  for Drell-Yan process, which  gives the distribution of initial partons. We have employed  the translation invariance and integrated on $x_1$, which gives $q=q'$.  

On the other hand,
\begin{eqnarray}
\label{dmunu}
D_{\mu \nu}&=&\int \frac{d^3K}{(2 \pi)^3 2E} \frac{d^3k'}{(2\pi)^3 2E'} \int d^4 x_2 d^4 x_3 d^4 x_5 e^{-ik_{\bar c}x_2} e^{ik_{\bar c}x_5} \sum \limits_{ X_h}\nonumber\\
&\times &(\gamma_\mu (\not k'+ m) \gamma_\nu)_{\alpha_5 \beta_2} <0|(\bar \psi(x_3) \not V(x_3))^{j_3}_{\beta_3}|X_h><X_h|(\not V(0) \psi(0))^{j_0}_{\alpha_0} |0> \nonumber\\
& \times &<0|\psi^{j_3}_{\beta_3}(x_3) \bar \Psi^{j}_{\alpha_5}(x_5)|\bar D><\bar D|\Psi^{j}_{\beta_2}(x_2) \bar \psi^{j_0}_{\alpha_0}(0)|0>.
\end{eqnarray}

To get the above expression, we have written the total final state produced  in the A B collision as $|\bar D X>= |\bar D \tilde{X} c X_h>$, where the $\tilde X$ represents all the particles except the $c~ \bar c$ produced by the hard interaction and $X_h$ for the assemble of particles  produced by all the interactions  except the above hard one.  We have the corresponding field acting  on the c quark final  state with momentum $k'$,  $k_{\bar c}=q-k'$. The summation on $\tilde{X}$  has been  eliminated  by the completeness condition.

In (\ref{jiemian2}), C is the colour factor of the partonic diagram except the external leg combined into $\bar D$. The colour part will be clarified following. 


The $W^{\mu \nu}$ can be conventionally written as
\begin{eqnarray}
\label{wmnfac}
\int \frac{d^4 q}{(2 \pi)^4} W^{\mu \nu}&= & \int dx_1 dx_2 \int \frac{d \lambda_1}{2 \pi} \frac{d \lambda_2}{2 \pi} e^{-ix_1 \lambda_1} e^{-i x_2 \lambda_2} <A| \bar \psi(y)\frac{\gamma^+}{2P^+_1}\psi(0)|A> \nonumber\\
&\times &  <B| tr(\frac{\gamma^-}{2P^-_2}\psi(y) \bar \psi(0))|B> tr(\frac{\not P_1}{2} \gamma^\mu \frac{\not P_2}{2} \gamma^\nu),
\end{eqnarray}

which is the intended factorized form.  
  Some of the above  variables are: $\lambda_1=P_1^+y_-, \lambda_2 =P_2^-y_+ $, $y=(y^+, y^-, {\bf 0}_\perp)$, $ Q^2 \equiv q^2 =x_1 x_2 s$, s is the CMS energy  for the nucleon-nucleon system whose partons collide and produce the heavy quark pair.

The factorization for $D_{\mu \nu}$ is more complicated. It  can be written as:

\begin{eqnarray}
\label{dfac3}
& & \int \frac{d^3 K}{(2 \pi)^3 2E} d z K^+ d z_l K^+ (2 \pi) \delta(k'^2-m^2) (2 \pi)^4 \delta^4(k_l + k_{\bar c} -K)\nonumber \\
&\times &tr(\gamma_\mu (\not k'+m) \gamma_\nu \frac{\not K}{2})|_{k'=q-zK^+} \sum \limits_{X_h} \nonumber\\
&\times &\int d^4 x_3 e^{-i k_l x_3}|_{k^+_l = z_l K^+} <0| (V(x_3) \bar \psi(x_3))_{j_0}|X_h><X_h| \frac{\gamma^+}{2 K^+}(\psi(0) V(0))_{j_0}|0>\nonumber \\
& \times &\frac{1}{9} \int \frac{d^4 k_{\bar c}}{(2\pi)^4} \frac{d^4 k_{l}}{(2\pi)^4} \delta(z-\frac{k^+_{\bar c}}{K^+}) \delta(z_l-\frac{k^+_{l}}{K^+}) \int d^4 x_2 d^4 x_5 e^{-i k_{\bar c} x_2} e^{i k_{l} x_5} \nonumber \\
& \times & <0| tr(\frac{\gamma^+}{2 K^+} \psi^{j_0}(x_5) \bar \Psi^{j}(0))|\bar D>
<\bar D| tr ( \frac{\gamma^+}{2 K^+} \Psi^{j}(x_2) \bar \psi^{j_0}(0))|0>
\end{eqnarray}

In the above equation, 
the colour indices in the partonic final states and the distribution functions  are summed. So the colour indices in the combination matrix elements should be averaged ($\frac{1}{9}$), which is similar  as the case in fragmentation function. We do not separate the colour-singlet or the colour-octet contribution in the matrix elements, but sum them together. The reason is that  the partonic cross section is the same for the colour indices $j$  belonging  to $\underline{1}$ or $\underline{8}$ states (since the other parton comes from an un-correlated source), and that for the parton distribution in QGM, it should be the same whether $j_0$ belong to  $\underline{1}$ or $\underline{8}$ states. We have taken the external field proportional to $n^\mu$, which select only the ``+'' component since $\not n  K \fslash ^- \not n = \not n  {\bf K}\fslash  _\perp \not n=0$.

In the part corresponding to the partonic sub-process, for the sake of factorization, we have done the collinear expansion for the momentum of the $\bar c$  along the ``+''  component of the momentum of $ \bar D$.  Here one notices  that the coordinate system is different from that for initial states $W^{\mu \nu}$. The z direction is along the the momentum of the anti-charm meson.

In Equation (\ref{dfac3}), we write the combination matrix element (Row 4, 5) formally to be analogous to that in other processes (see Section \ref{sec:come}).  $z$ and  $z_l$ seem not restricted to be $z+z_l=1$. However,   the $\delta$ functions in the first row sets  the restriction. At the same time, the integral in the combination matrix elements $\int{d^4 k_l}$ acts on the matrix element corresponding to the quark distribution in the QGM (Row 3)  as well as the $\delta$ function in the first row. These show that we have not finished the factorization. To get the factorized form, we notice

\begin{eqnarray}
&~ & \int \frac{d^3 K}{2E}\delta^4(k_l + k_{\bar c} -K) \nonumber \\
&= & \int \frac{dK^- d^2K_{\perp}}{2K^-}\delta^+( k^+_l + k^+_{\bar c} -K^+)\delta^- \delta^2 _\perp \nonumber \\
&= & \int \frac{d^3 K}{2E} 2E \delta^3 ({\bf k}_{\bar c}+{\bf k}_l -{\bf K}) \frac{1}{2K^-}\delta^+( k^+_l + k^+_{\bar c} -K^+).
\end{eqnarray}

Let the 3-dimension $\delta$ function  absorbed into the combination matrix element, we get the `restricted' matrix element or the dimensionless combination function: 
\begin{eqnarray}
 \tilde{F}(z, z_l)&=&\frac{1}{9} \int \frac{d^4 k_{\bar c}}{(2\pi)^4} \delta(z-\frac{k^+_{\bar c}}{K^+})\frac{d^4 k_l}{(2\pi)^4} \delta(z_l-\frac{k^+_{l}}{K^+})  \int d^4 x_2 d^4 x_5 e^{-i k_{\bar c} x_2} e^{i k_l x_5} \nonumber \\
&\times & <0| tr(\frac{\gamma^+}{2 K^+} \psi^{j_0}(x_5) \bar \Psi^{j}(0))|\bar D>
<\bar D| tr ( \frac{\gamma^+}{2 K^+} \Psi^{j}(x_2) \bar \psi^{j_0}(0))|0> \nonumber \\
&\times & 2 E \delta^3 ({\bf k}_{\bar c}+{\bf k}_l -{\bf K}).
\end{eqnarray}

The integral of $z_l$ in the first row of Equation (\ref{dfac3}) have given $z+z_l=1$.
$\int dz  \delta(k'^2-m^2)$  gives important result:  $z=\frac{Q^2}{2K\cdot q}$, \footnote{This relation is exact for $Q$ and $K$ to be infinite while $z$ fixed. It is a good approximation when  the anti-quark to be combined into the heavy meson  is on mass shell in the partonic processes. In this case,  the four momentum of the anti-quark is $(zK^+, \frac{m^2}{2zK^+}, {\bf 0}_\perp)$. In the heavy meson rest frame,  it is easy to get $z =\frac{\sqrt{m^2+(mv)^2}+mv}{M} \simeq \frac{m}{M} $. Go back to the initial parton CMS, use this approximation for the on-shell condition  $(q-k)^2=m^2$, we can get the relation. So, it is not just an approximation by taking $M=m_c=0$ in $K \cdot q$.    }  which is analogous to the Bjorken scaling variable. This means that for certain partonic CMS energy $Q$, the heavy meson with momentum $K$ just comes from the heavy quark with momentum fraction $z$ combined with light quark with momentum  $k_l=(1-z)K$, hence only probe this light quark in QGM.  Such a conclusion does not depend on the special forms of the derivation in this note. In fact, the relation $z=\frac{Q^2}{2K\cdot q}$ is set by the on shell condition of the heavy quark   associatively produced  with the one to be   combined into the final state heavy hadron. Just like the DIS process, this is a physical condition which should be respected by any  special forms  of derivation. 

The  cross section section now can be written as   



\begin{eqnarray}
\label{cs11m}
2 E \frac{d \sigma_C}{d^3 K}&=& \frac{1}{4 F} \sum \limits_{ab} \int{dx_1 dx_2} 2 f^a_A(x_1) 2 f^b_B(x_2) \nonumber \\ 
& \times & | \tilde{\cal M}_{ab} | ^2  \frac{1}{z} \frac{(2 \pi)^2}{(2M)^2}  2 P (z_l) \tilde{F}(z, z_l) | _{z+z_l=1}
\end{eqnarray}


Here $f^a_A$ and $f^b_B$ are parton distributions in Nuclei and should be treated by the Glauber geometrical formulae \cite{glauber}. $ M$ is the mass of the heavy meson.  $| \tilde{\cal M}| ^2$ refers to the invariant amplitude square including all the coupling constant and colour factors for the partonic process $ab \to \bar c +x$ (where the momenta of external legs are modified and $x$ to be considered as one particle). For example,   for $q \bar q \to \bar c +x$, to the lowest order,  $| \tilde{\cal M}| ^2$  is
\begin{equation}
\label{partoniccs}
(4 \pi \alpha_s)^2 C' tr(\frac{ P \fslash _1}{2} \gamma^\mu \frac{ P\fslash _2}{2} \gamma^\nu) \frac{1}{q^4} tr(\gamma_\mu (\not k'+m) \gamma_\nu \frac{ K \fslash }{2}). 
\end{equation}

$C'$ is the colour factor. Though the quark mass term is vanishing, we keep it to show the origin of the formula.

$P(z_l)$ can be understood as the distribution function of the parton (probed by the heavy quark) in the external source.  $P(z_l)$ is also dimensionless: 

\begin{equation}
\label{litd}
\frac{1}{2} \int{d^4 x_3} e^{-ik_lx_3} | _ {k_l^+=z_l K^+} \sum \limits_{X_h} <X_h| tr (\psi(0) V(0)|0> <0|V(x_3)\bar \psi(x_3) \frac{\gamma^+}{2 K^+})|X_h>
\end{equation}

This can be understood as the expectation value on the  state representing an assemble of  particles produced in the A B collision denoted by $|X_h>$. 



For numerical calculation,  we give more discussions on the cross section formula.
The heavy quark production sub-process in nucleus-nucleus collision is hard interaction and can be calculated in the binary approximation. The initial state for this sub-process  in Equation (\ref{jiemian1}) then can be understood as nucleus modified nucleon.  We use the Glauber geometrical formulae \cite{glauber} to describe the distribution of nucleon in the nucleus. i.e., 
for a certain impact parameter $b$, the production (interaction) number ${\cal N}$ is

\begin{equation}
\label{nuclearintnum}
\frac{d{\cal N}_C}{dR}= T_{AB}(b)\frac{d\sigma_C^{\tilde{N} \tilde{N}}}{dR}.
\end{equation}

In the above equation, $T_{AB}(b)$ is the overlap function for nucleus A and B \cite{glauber}.
$\frac{d\sigma_C^{\tilde{N} \tilde{N}}}{dR}$ is differential cross section for nucleon nucleon interaction calculated by Equation (\ref{cs11m}) with the initial state $|A B> \to |\tilde{N} \tilde{N}>$. $dR$ is phase space element for the final state. The  `$\tilde{~}$' on $N$ indicates that  the nucleon is  modified in nucleus. Now In Equation (\ref{cs11m}), the parton distribution function is for partons in modified nucleon, The incident flux factor in Equation (\ref{jiemian1}) is for nucleons and can be absorbed into the partonic cross section same as in the case of Drell-Yan process, with the above trace term (\ref{partoniccs})  slightly modified to be the exactly partonic invariant amplitude square.  




Now the production number of $\bar{D}$  in the combination process is 
\begin{equation}
\label{cossnum}
2 E \frac{d {\cal N_C}}{d^3 K}= T_{AB}(b)\sum \limits_{ab} \int{dx_1 dx_2}  f^a_1(x_1)  f^b_2(x_2)
\frac{d \hat{\sigma}_{ab}}{d {\cal I}} \frac{1}{z^2} \frac{(2\pi)^2}{(2M)^2}  P(z_l) \tilde{F}(z, z_l)| _{z+z_l=1}.
 \end{equation}
To get results for our relevance, we can integrate over $b$ in central region. In the above equation, 
$d{\cal I}$ is the dimensionless invariant phase space for the `2-body' partonic final state  $\bar c + x$ where $x$  treated as   one particle. 
This formula is also correct for higher order partonic cross sections. 


 
\section{The universal (process-independent) combination  matrix elements}
\label{sec:come}

From the  above section, The cross section of the $\bar{D}$ is  dependent on both the combination matrix elements and the distribution of the light quark in the QGM. If we compare with  data to extract the light quark distribution $P(z_l)$, one of the  key inputs is the combination function  $\tilde{F}(z, z_l)$, which is not calculable by PQCD and we should find ways to extract from more `simple'  experiments. This require that the cross section of a  more simple process can be factorized and includes this parameter. The following is an example. Let's see the factorization and the complexity.

It has been pointed out that, in hadronic interaction, the asymmetry of D meson in forward direction can be explained by the the combination of the initial parton with the charm quark produced in the hard interaction. Such a  leading particle effect  has been studied in  \cite{lpef}, \cite{masi}, both in the approximation $m_c \to \infty$. In such an approximation, the light quark has vanishing momentum, hence, qualitatively, the momentum of the D meson is approximately that of the charm quark, so that   not possible to probe the momentum of the light quark. On the other hand, in \cite{masi}, the authors also  tried to give the combination matrix elements in the framework of collinear factorization, which is the same framework used in this note.  The combination matrix elements  there depend on  3 variables $z_1$, $z_2$, $z_3$, seem not corresponding to the momentum fraction of the valence partons. 
However, starting  from Equation (4) in \cite{masi}, by taking into account the space-time transition invariance, we get the combination matrix elements with 2 variables corresponding to the momentum fraction of the charm and the light quarks, which is like those in the above section:
\begin{eqnarray}
\label{uccomme}
& & \int \frac{d^4 k_{c}}{(2 \pi)^4} \frac{d^4 k_{l}}{(2\pi)^4} \delta(z-\frac{k^+_{c}}{K^+}) \delta(z_l-\frac{k^+_{l}}{K^+}) \int d^4 x_1 d^4 x_2 e^{i k_{c} x_1} e^{-i k_{l} x_2} \nonumber \\
&\times  & <0| \bar q_{k}(0) \frac{\gamma^+}{2 K^+} Q_{l}(x_1))|H_Q>
<H_Q| \bar Q_{i}(0) \frac{\gamma^+}{2 K^+}  q_j(x_2))|0>,
\end{eqnarray}

and

\begin{eqnarray}
\label{uccomme2}
& &\int \frac{d^4 k_{c}}{(2 \pi)^4} \frac{d^4 k_{l}}{(2\pi)^4} \delta(z-\frac{k^+_{c}}{K^+}) \delta(z_l-\frac{k^+_{l}}{K^+}) \int d^4 x_1 d^4 x_2 e^{i k_{c} x_1} e^{-i k_{l} x_2} \nonumber\\
&\times  & <0| \bar q_{k}(0) \frac{\gamma^5 \gamma^+}{2 K^+} Q_{l}(x_1))|H_Q>
<H_Q| \bar Q_{i}(0) \frac{\gamma^5 \gamma^+}{2 K^+}  q_j(x_2))|0>.
\end{eqnarray}
\

The two parts of the combination matrix element, i.e., the double-vector part and the  double-pseudo-vector part, should be separated here since the partonic cross sections corresponding to these two parts could be different. For the case that the  quark and the anti-quark  from different sources respectively, as  in the process in Section \ref{sec:formucom}, these two parts can be put together and only the vector part needs  consideration.

The complexity lies in that,  Equations (\ref{uccomme}, \ref{uccomme2}) are different from that in Section \ref{sec:formucom} by  $2 E \delta^3 ({\bf k}_{\bar c}+{\bf k}_l -{\bf K})$,  without the restriction $z+z_l=1$.  The reason is that in the process in \cite{masi},  the  light quark and the  heavy quark can undergo hard interactions and in principle are not restricted on mass shell.  We can also understand this from a different way. Equations (\ref{uccomme}) and (\ref{uccomme2})  look like the combined distribution of two valence quarks in the heavy meson. In fact, rewriting the $<0|(\cdot \cdot \cdot)_1 |H_Q><H_Q|(\cdot \cdot \cdot)_2|0> $ to the form $<H_Q|(\cdot \cdot \cdot)_2 |0><0|(\cdot \cdot \cdot)_1|H_Q> $, integrating the $\delta$ functions and the exponential functions, taking $|0><0|=1$ in the vacuum saturation approximation, we will get the form of the product of two parton distribution functions, each similar to that defined by  Collins and Soper \cite{collinsoper}. Then in a parton model at high energy, we can not require the sum of two parton momentum fractions equals one. Hence to get the inputs needed, we should find ways to relate the `restricted' in Section \ref{sec:formucom} and the `unrestricted' ones here.

If the above matrix elements have been  extracted from experiments, 
to get  the restricted  combination function, we start from, by denoting the Combination Matrix Element in Equations (\ref{uccomme},\ref{uccomme2})  as $CME$:

\begin{eqnarray}
\label{CMEeq}
CME&=&\int \frac{d^3 K'}{2E'} 2E' \delta^3({\bf k_c'+k_l'-K'}) \frac{d^3 k_c'}{2E_c'} 2E_c' \delta^3({\bf k_c'-k_c}) \nonumber \\ 
&\times & \frac{dk_l'}{2E'_l} 2E_l' \delta^3({\bf k_l'-k_l})  CME \nonumber
\\
&=& \int \frac{d^3 K'}{2E'}  \tilde{F}(z,z_l; z+z_l)
\end{eqnarray}

In principle, we should solve the integral equation and use the value of $\tilde{F}$ on $z+z_l=1$ (${\bf K'=K}$) as our inputs.  On the other hand, if the real world is more simple --- as most models assume,   $\tilde{F}(z,z_l; z+z_l)$ peaks around $z+z_l=1$, i.e.,  two valence quarks on mass shell with ${\bf k_c+k_l \simeq K}$ 
--- we can, approximately, fit  $CME$ as,

\begin{eqnarray}
CME&=&\int \frac{d^3 K'}{2E'} 2E'\prod \limits_{i}\frac{\epsilon_i}{\pi (\epsilon_i^2+(K-K')_i^2)} M^2 f(z,z_l) \nonumber\\
(\epsilon_i \to 0) &=& \int \frac{d^3 K'}{2E'} 2E'\delta^3({\bf K}- {\bf K}') M^2 f(z,z_l)
\end{eqnarray}

So in the extreme/ideal condition, we can just have $\tilde{F}(z,z_l)_{z+z_l=1} \propto \frac{CME}{M^2}$. That is, because  the distribution of $\tilde{F}$ is a narrow peak, we use the average value of it in a reasonably small integral region of $\frac{d^3 K'}{2 E'}$.


Equations (\ref{uccomme}, \ref{uccomme2}) will be applied in other processes and discussed in elsewhere, so that we can have more experiments to extract the combination matrix elements. In principle, when we accumulate enough number of data, especially from more than one process, the integral equation (\ref{CMEeq}) can be solved. We just mention that we have discussed the combination process preliminarily  in $ e^+$  $e^-$ annihilation, where a light quark `fragments' into a heavy meson by combination \cite{jinli}.

\section{Numerical estimation}
\label{sec:num}

%
Since there are yet  no data  on the combination matrix elements and  the heavy flavour particle  production in heavy ion collisions,  
only to get a practical view of the discussions in this note,  we give a numerical estimation of the transverse momentum distribution of the open charm mesons  produced via combination by assuming   the  forms of the combination matrix elements and the parton distributions of the `QGP'. 

We make a simple assumption that the combination matrix elements  are not sensitive to  $ z$ and $z_l$ and just take them as constant around $z + z_l =1$. At the same time, because of no data, we can not specify any identified D meson, whatever $D^0$, $D^{\pm}$ or $D^{*}$.
 
There are many discussions  about the possible distributions of the partons in the hot dense quark-gluon matter produced in RHIC. Especially, in all the  combination models cited above, this is one of the key inputs. It is generally adopted that  the distribution function  of the thermal partons are exponential while those produced  from  the initial hard interactions  are of minus powers. In the following we use some minus power, exponential and Gaussian distributions to see their differences.

Because the absolute values of the parameters are unknown, as a demonstration of the combination effects, we  study the ratio of the $\bar D$ spectrum w.r.t. the charm quark  spectrum, $\frac{d {\cal N}^{\bar D}}{ dp_\perp} / \frac{d {\cal N}^c}{ dp_\perp}$ 
and normalize the total ratio (integrated on $p_\perp$) to 1. 

Besides, there still some other things to be stated:

1) The initial parton distribution function of nucleon in nucleus. Now there are several groups of nPDF's \cite{npdf}. 
In this note, we just use the EKS one available in CERN PDF Library (version8.04) \cite{pdflib}.

2)  
  For simplicity, we  use the LO partonic cross section. It is clear that to get the partial cross section of combination process from data,  the contribution by heavy quark fragmentation should be calculated and subtracted from data. 
 Such a  work needs the NLO partonic cross section (see following discussion on the RHIC d Au data) and is  in progress. 
   
3)  After the heavy quark created, it inevitably go through the QGM and loses energy by radiation. 
From  Equation (\ref{cossnum}), we can see that  $P(z_l) \bigotimes \tilde{F}(z, z_l)$ play the r\^{o}le of Fragmentation Function (FF) of the heavy quark in this process and the radiation of gluons can be treated as the evolution of this FF (similar as the FF of a quark into the photon, see \cite{lwp}). This leads to that, FF is originally function of $z$, now also of  the scale $\mu^2$.  In other words, the assumptions on $P(z_l)$ and $\tilde{F}(z, z_l)$ should be understood as the values on $\mu^2=Q^2$.

\begin{figure}
\psfig{file=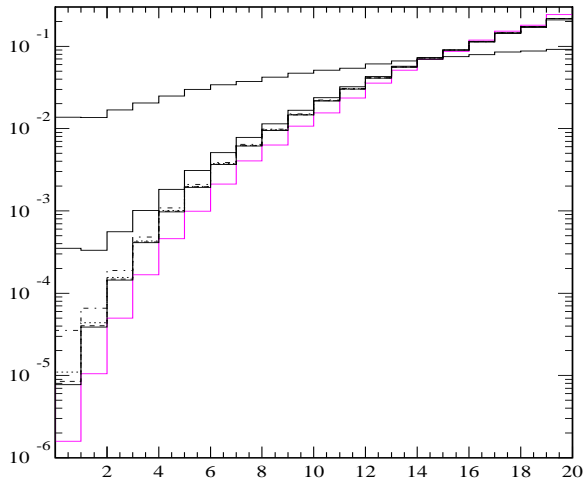,height=8cm,width=10cm}
\caption{The ratio of the spectrum of $\bar D$ meson to that of  charm quark, with  total ratio normalized to  1. At the most left: the highest curve  is for that the parton distribution of QGM, $P(z_l)$, is exponential; the lowest for $P(z_l)$ is Gaussian; the ones between them for  $P(z_l)$'s  are of various powers. }
\label{ally12}
\end{figure}

\begin{figure}
\psfig{file=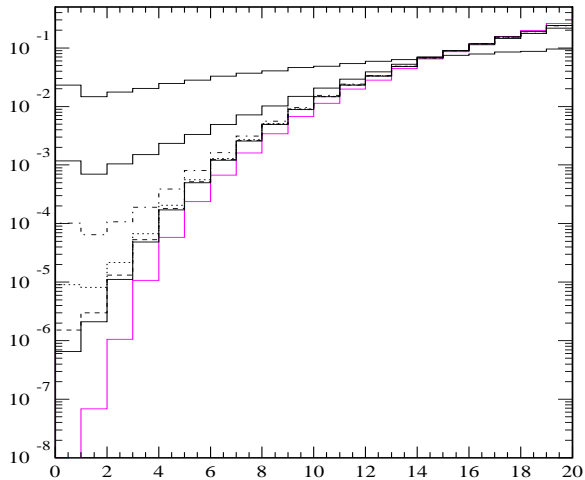,height=8cm,width=10cm}
\caption{Same as Figure (\ref{ally12}), except that the rapidity of the particles are restricted to central region ($|y|<0.5$). }
\label{cony12}
\end{figure}

Figures  (\ref{ally12}) and (\ref{cony12}) give the numerical results.  

From the most simple observation, because the heavy meson comes from the heavy quark combined with  a light one,  the momentum of the meson are always  larger than  that of the heavy quark. Such a fact is shown in the  Figures that all the spectra are harder than that of the $c $ quark. 
The curves are sensitive to the distributions, so that we can measure the parton distribution of the QGM from the heavy meson spectrum.


Experiments  are more interested  on central rapidity region. We also  give the ratio in the region $|y|<0.5$. The curves have the same property of those in the whole rapidity region. 


The RHIC data of D mesons in d Au collision  show that their  spectrum  is as hard as that of  the charm quark calculated at NLO. 
This may be an indication of combination  (see  \cite{huang}). Of course  the external source for the light quark in this case is  cold quark matter.  
Our calculation is consistent with these data qualitatively.
The  reason that we can not give any quantitative prediction or explanation  is that the combination matrix elements are yet beyond approach. 

%

\section{Conclusions and discussions}
\label{sec:concl}


The discovery of the extreme state of matter such as QGP  is not the end of the story. We should continue to look for more  ways  to measure it in details. The combination process of heavy hadron production discussed in this note  may shed light on such a purpose.

The `combination rules', suggested in various forms in different combination models,
are  defined by  universal (process-independent)  matrix elements in this note. They are model independent and   can be extracted from other processes than the complex relativistic heavy ion collision. 
In  the same framework,  the `parton distribution function'  of QGM  (if allowed to be given such a `standard' name),  here also has an operator definition and is model independent. 

From the cross section formulae in this note, we can see that the matrix elements have the `$z$ scaling behaviour'. We also have argued that the scaling violation can be explored by taking into  account the  radiation correction.

The result of this note is just the  beginning of a series of   systematic works to be done, in theory as well as in experiment. This is just like the case in  studying the parton distribution functions and fragmentation functions of hadrons.
Among the works,  the extraction of the combination matrix elements  from  various processes is basic and crucial for understanding  the forthcoming experimental data.


 

One thing  many experimentalists and theorists are concerned is  the collective behaviour of heavy quarks. The combination formulae above provide a baseline for observing this issue.  If some collective behaviour is measured on final state heavy hadrons, and if  it only comes from the light quark which the heavy quark combined with,  the value of the parameter, e.g., the $v_2$ of the heavy meson should be $\frac{1}{2}$ or $\frac{1}{3}$ times of light meson or light baryon, respectively, which is the value for a single quark. Only if the experimental  value is larger than that, we can expect that the heavy quarks also have collective behaviour. This is a very interesting topic in study of the vacuum structure. In the $T=0$ vacuum, the heavy quarks can hardly be produced via soft interactions, e.g., `tunneling effects'. However,  if the temperature is extremely high and the QCD vacuum is changed, the heavy quark pair may be created and their collective effects may reflect
the properties of the bulk.

The author thanks Prof. X.-N. Wang for suggestion of studying this topic, and helpful discussions. He  also thanks members of the Theoretical Particle Physics Group of Shandong University for encouraging and helpful discussions, especially  Prof. Z.-G. Si for the discussion on combination matrix elements for the leading particle effects.
This work is supported in part by the National Natural Science Foundation of China (NSFC) under Grant  10205009.


\begin{thebibliography}{99}
\bibitem{preqgp}
 This was predicted soon after QCD was established and the asymptotic freedom of which was discovered, see, 
 J. C. Collins, M. J. Perry, Phys. Rev. Lett. 34: 1353, 1975.
For list of historical papers, see the references in \cite{jacobwang}.


\bibitem{signal}
 PHENIX Collaboration, K. Adcox, et al, nucl-ex/0410003;  

 STAR Collaboration, J. Adams, et al, nucl-ex/0501009.
                                                                                


\bibitem{nature}
G. Brumfiel, Nature 430, 498 (29 Jul 2004).

\bibitem{jacobwang}
 Review and discussion from the theoretical aspect, see, e.g., P. Jacobs, X.-N. Wang, hep-ph/0405125.



\bibitem{fries}
R. J. Fries, B. Muller, C. Nonaka, S. A. Bass, Phys. Rev. Lett. 90: 202303, 2003.

\bibitem{hwa}
R. C. Hwa, C. B. Yang, Phys. Rev. C67: 034902, 2003.

\bibitem{ko}
V. Greco, C. M. Ko, P. Levai, Phys. Rev. Lett. 90: 202302, 2003.

\bibitem{annisovich73el}
V. V. Anisovich, V. M. Shekhter, Nucl. Phys. B55: 455, 1973;

J. D. Bjorken, G. R. Farrar, Phys. Rev. D9: 1449,  1974;

 K. P. Das, R. C. Hwa, Phys. Lett. B68: 459,1977, Erratum-ibid. B73: 504, 1978;

 E. L. Berger, T. Gottschalk, D.  W. Sivers, Phys. Rev. D23: 99, 1981;


 Qu-Bing Xie, Xi-Ming Liu, Phys. Rev. D38: 2169, 1988.

 Combination/Coalescence models in heavy ion collisions, see:

 P. Koch, B. Muller, J. Rafelski, Phys. Rept. 142: 167, 1986;

 J. Rafelski, M. Danos, Phys. Lett. B192: 432, 1987.
  

\bibitem{xieshao}
 Feng-lan Shao, Qu-bing Xie, Qun Wang, nucl-th/0409018.

\bibitem{NLOte}
The most to-date comparison between experiment and theory, see:
 
CDF Collaboration,  D. Acosta, et al, hep-ex/0412071;

  M. Cacciari, S. Frixione, M. L. Mangano, P. Nason, G. Ridolfi, JHEP 0407: 033,  2004.

\bibitem{dong}
 R. Baier, Yu. L. Dokshitzer, S. Peign\'{e}, D. Schiff, Phys. Lett. B345: 277, 1995;
 
M. Gyulassy, P. Levai, I. Vitev, Phys. Rev. Lett. 85: 5535, 2000;

 M. Gyulassy, P. Levai, I. Vitev, Nucl. Phys. B594: 371, 2001;
 
U. A. Wiedemann,  Nucl. Phys. B588: 303, 2000;
 
X.-F.  Guo, X.-N. Wang, Phys. Rev. Lett. 85: 3591, 2000;

X.-N. Wang,  X.-F.  Guo,  Nucl. Phys. A696: 788, 2001.
 
 
 
\bibitem{wanggy}
 M. Gyulassy,  X.-N. Wang, Nucl. Phys. B420: 583, 1994;

 X.-N. Wang,  M.  Gyulassy,  M.  Plumer, Phys. Rev. D51: 3436, 1995.


\bibitem{qiuvitev}
 J.-W.  Qiu, I. Vitev, Phys. Lett. B570: 161, 2003.


\bibitem{glauber}
K.J. Eskola, K. Kajantie, J. Lindfors,  Nucl. Phys. B323: 37, 1989; 

R. J. Glauber, in {\it Lectures in Theoretical Physics}, Eds., W. E. Brittin, L. G. Dunham (Interscience, NY, 1959), Vol. 1, p. 315.

\bibitem{lpef}
E.  Braaten, Y.  Jia,  T.  Mehen,  Phys. Rev.  Lett. 89: 122002, 2002; Phys. Rev.   D66:  034003, 2002; Phys. Rev. D66:  014003,  2002.


\bibitem{masi}
C.-H.  Chang, J.-P.  Ma, Z.-G. Si,  Phys. Rev. D68: 014018, 2003.

\bibitem{collinsoper}
J. C. Collins, D. E. Soper, Nucl. Phys. B194, 445:1982.

\bibitem{jinli}
JIN Y.,  LI S.-Y., XIE Q.-B., High  Ener. Phys. Nucl. Phys., 27: 852, 2003   
(in Chinese).

\bibitem{npdf}
K.J. Eskola, V.J. Kolhinen, C.A. Salgado, Eur. Phys. J. C9: 61, 1999;
Nucl. Phys. B535: 351, 1998;

M. Hirai, S. Kumano, M. Miyama, Phys. Rev. D64: 034003, 2001;

D. de Florian, R. Sassot, Phys. Rev. D69: 074028, 2004.

\bibitem{pdflib}
Plothow-Besch, Int. J. Mod. Phys. A10: 2901, 1995.

\bibitem{lwp}
S.-Y. Li et al, in preparation.

\bibitem{huang}
see, H. Z. Huang, J. Rafelski, hep-ph/0501187.



\end{thebibliography}
\end{document}